# Predicting Students' Learning Styles Using Regression Techniques

*Mohammad Azzeh, Ahmad Mousa Altamimi, Mahmoud Albashayreh*
*Department of Computer Science*
[1]*Department of Data Science, Princess Sumaya University for Technology*
[2,3]*Applied Science Private University, Amman, Jordan*
*m.azzeh@psut.edu.jo, a_altamimi@asu.edu.jo, m_albashayreh@asu.edu.jo*

Abstract- Traditional learning systems have responded quickly to the current circumstances due to the Covide-19 pandemic and moved to online or distance learning. Online learning ensures continuity of education, enriches the conventional learning models, and promotes active learning. In its application, online learning requires a personalization method because the interaction between learners and instructors is minimal, and learners have a specific learning method that works best for them. One of the personalization methods is detecting the learners' learning style as learners have their own learning preferences. To detect learning styles, several works have been proposed using classification techniques. However, the current detection models become ineffective when learners have no dominant style or have a mix of learning styles. Thus, the objective of this study is twofold. Firstly, constructing a prediction model based on regression analysis to provide a probabilistic approach for inferring the preferred learning style. Secondly, comparing between regression models and classification models for detecting learning style. To ground our conceptual model, a set of machine learning algorithms have been implemented based on a dataset collected from a sample of 72 students using VARK's[1] inventory questionnaire. A student can answer each question of the questionnaire by assigning one or more style from a set of four learning styles. Results show that regression techniques are more accurate and representative for real-world scenarios than traditional classification algorithms where students might have multiple learning styles but with different probabilities. The accuracy results and significance tests confirm that Random Forest is the most superior regression method. We believe that this research will help educational institutes to understand and engage learning styles in the teaching process to maximize student learning outcomes.

1. Introduction

Due to the Covid pandemic, the largest disruption to education in history has been recorded, which has had a nearly universal impact on learners and educators worldwide [1]. As a result, learning systems have successfully undergone critical changes and moved to use new models (i.e., online learning or distance learning) supported by information and communication technologies [2]. Online learning enriches conventional learning by offering flexibility and self-paced learning with an efficient way to deliver knowledge through virtual communication and collaboration [3]. However, it requires a personalization method as learners have different backgrounds, knowledge, and various learning environments [4]. One of the personalization methods is detecting the learners' learning style as it influences individual academic achievement [5]. The concept of learning style has been coined during the mid of 70's and is formally defined as: "an individual's mode of gaining knowledge" [6]. In other words, it is the best method a person uses to learn.
Many learning styles' theories have been introduced in the field of education and widespread recognition in education theory and learning strategies [7]. These theories include ones that classify people according to their own distinguishing features that differentiate one from others [8]. The widespread of theories is motivated by the fact that knowing a learner's learning style can enable instructors to maximize learners learning by using adapted teaching methods and allow them to recognize their learning styles to find what study methods and activities help them learn best [9]. Thus, the awareness of the learning style roles in the education process is very important for both learners and researchers [10].

---

[1] VARK stands for (Visual, Auditory, Reading/writing, and Kinesthetic)

To recognize individuals learning styles, inventories can be used, typically take the form of a questionnaire assessment, where a series of questions are asked and then scored the results to illustrate the dominant learning styles [11]. There are many popular learning style inventories proposed in the literature, such as Fleming's VARK[2] Learning Style Questionnaire [12], Kolb's Learning Style Inventory (LSI) [13], Jackson's Learning Styles Profiler (LSP) [14], and other. Each of these proposed a set of questions to identify the learners' different styles. For example, according to VARK, learners are categorized into four different types: Visual, Auditory, Reading/writing, and Kinesthetic [12]. On the other hand, Kolb's, which is also one of the most widely used inventory, identified four learning styles: Diverging, Assimilating, Converging, and Accommodating [13].

More recently, a considerable amount of research has been devoted to automatically detect the learning style [15][16][17]. The main approach was based on machine learning techniques, where classification and clustering algorithms have been applied. Being said that, classification algorithms was the dominate of these works. For example, Decision Tree was used in [18] to detect the learners' learning styles from students' weblogs. The authors of [19] also used the Decision Tree C4.5 algorithm to identify the learning styles from a sample of 1,205 students collected using the VARK questionnaire. Neural Networks was also employed in [20]. On the other hand, Other clustering techniques have also been utilized such as Fuzzy C-Means was employed in [16] as a clustering algorithm to detect learner's styles. The authors of [17] utilized K-modes clustering algorithm to improve the e-learning system. The model was implemented based on the Felder and Silverman learning style model and used a real dataset extracted from an e-learning system's log file that captures the learners' behaviors.

As one can be noticed, almost all of the proposed works were designed to detect the learners' learning styles and identified a single learning style for each learner. But in practice, learners might have a single or multiple learning styles, where one can equally prefer both visual and auditory learning styles. For learners with a mix of learning styles (with probability) or with no dominant style of learning, detecting a single learning style is becoming ineffective. This was supported by the work presented in [21], where the researchers proved that learners have different learning styles, and thus, no single system can serve well with all learners. Therefore, the current approaches do not support this trend, and thus a new learning style detection system has been proposed to solve this issue based on regression analysis. According to the best of our knowledge, no work has provided a probabilistic approach to infer the learner's style. Hence, this work utilizes regression analysis to provide a probabilistic approach for inferring the preferred learning styles.

To this end, a dataset was collected using the VARK's inventory questionnaire from a sample of 72 students. To easily collect the students' responses, we developed an online version of the questionnaire of 16 different questions using Microsoft Forms (part of Office 365). The responses were then imported as an Excel file and preprocessed to be eligible for the machine learning algorithm. Here, we divided the whole dataset into an array of four matrices (A, V, K, R), a matrix for each learning style. Each matrix consists of 16 columns demonstrating the presence or absence of a learning style and 5 output columns (4 columns represent the learning styles' probabilities which will be served as output for regression models, and the last column represents the actually selected learning style label which served as output for classification models). After preparing the dataset, multiple prediction models are developed using five machine learning algorithms (Multi-Layers Perceptron Neural Network (NN), Support Vector Machine (SVM), Decision Tree (DT), Random Forest (RF), and K-Nearest Neighbors). The constructed model attempts to predict the probability of each learning style to identify the most favoured styles. In this case the output of prediction would be in this format: <A=0.3, V=0.22, K=0.08, R=0.4>. Then a threshold can be specified to select the most favoured learning styles. To accomplish that, we compute the distance between the top learning style and remaining learning styles, then any learning style that falls within the distance given by the threshold is selected as nominated learning style. In the example above, if the threshold equals 0.2, then the selected learning style is {R, A, V}. If the threshold is 0.1, then the selected learning style set is {R, A}. We recommend the threshold value to be not very small, ignoring some interesting learning style or too large that involve all learning styles.

---

[2] (V: Visual, A: Auditory, R: Reading/writing, and K: Kinesthetic)

To compare our finding with the traditional classification approaches, the same set of machine learning algorithms is used to develop classification models to predict the learning style label. The purpose of this comparison is to examine the effectiveness of both approaches for learning style detection. One can observe from the results that none of the classification models is predictive with high accuracy. Moreover, the overall results are not encouraging, suggesting that none of the models can produce highly accurate predictions. So, one can conclude that regression algorithms are more accurate and representative for predicting learning styles' probabilities. All results have been statistically tested using Wilcoxon significance tests between each pair of models.

The remainder of this paper is organized as follows. The related work is introduced in Section 2, while the background material related to our approach is given in Section 3. Section 4 demonstrates the research methodology. The experimental work, along with the evaluation measures, is then given in Section 5. The discussion about our results is presented in Sections 6. Finally, Section 7 presents the conclusion and directions for future research.

## 2. Related Work

Many works have been proposed to detect the learning style in the literature [16][17] [18] [19]. However, data mining classification algorithms have dominant and are classified into two approaches: clustering and classification. For example, the authors of [17] utilized the K-modes clustering algorithm to improve the e-learning system. The model was implemented based on the Felder and Silverman learning style model and used a real dataset extracted from an e-learning system's log file that captures the learners' behaviors.

Other classification algorithms have also been used. The Decision Tree was used in [18] to detect the learners' learning styles from students' weblogs. The authors of [19] also used the Decision Tree C4.5 algorithm to identify the learning styles. Here, the sample was collected from 1,205 students using the VARK questionnaire. Other algorithms have also been utilized. The Neural Networks was employed in [20], where Felder-Silverman's model was used to identify four dimensions of learning styles. These dimensions are sensing or intuitive, active, or reflective, visual, or verbal, and sequential or global. Felder-Silverman's model was also used in [16], but the Fuzzy C-Means was employed as a clustering algorithm to detect learner's styles based on their data stored in the log files.

On the other hand, Genetic algorithms were employed to describe learning styles. The work in [22] defines a group of chromosomes and assigns the learner's action to each gene. Then used these genes generate new populations of chromosomes that describe learning styles. In the same vein, the work of [23] classified learners based on their learning styles by combining genetic algorithms with K-NN. In this work, the learners' behaviors are represented in an n-dimensional space. Learners are then considered to have the same learning style if they have a shorter distance to others.

Recently, educational data mining has been extensively considered in the literature. The educational data mining community defines it as an emerging discipline concerned with developing methods for exploring unique educational data types for better understanding settings the students learn in [24]. The spreading of educational data mining is due to the emergence of numerous public data mining tools such as R, WEKA, RapidMiner, and KNIME [25]. The work presented in [26] demonstrated a comparison between these tools, and it concludes that each of these tools has its advantages and disadvantages. Ultimately, we reiterate that the clustering algorithms are used to group learners and discover behavior patterns such as learning styles. Thus, this technique is adopted in our study.

## 3. Preliminary Material

### 3.1. Learning Style

Most students have a preferred way to learn and take in information in different ways. Some learn best by observing, some by listening, while others have to do it to learn it. This concept is called the learning style. A learning style is the best method a person uses to learn. Formally, it is "an individual's mode of gaining knowledge" [6]. Knowing a student's learning style can enable teachers to use teaching methods that maximize student learning and help the students. Students can recognize their individual learning styles to determine what study methods and activities help them learn best [9]. Thus, the awareness of the learning styles' roles in the education process is very important for both learners and researchers [10].

Research on learning styles proves that learners have different learning styles, and thus, no single system can serve well with all learners. In fact, the researchers found that there are seven different main types of learning styles [27]. These styles are explained briefly below:

- Visual Learning: Visual learners learn better when using pictures, images, symbols, and graphs when they are reading. Looking at charts and diagrams is preferred when taking their notes.
- Verbal Learning: Verbal Learners prefer using words, both in speech and writing. They are reciting information, writing notes, and highlighting the key points in their notes.
- Aural Learning: Technique involving the use of sound and music for understanding and memory. They prefer hearing or listening to easy spoken instructions rather than reading them.
- Physical Learning: The use of body, hands, and sense of touch are preferred by physical learners to learn. Their learning technique involves doing, touching, and building.
- Logical Learning: Logical learners are natural thinkers and often learn by asking many questions to understand the whole picture. They infer the patterns and relationships by thinking abstractly to classify and categorize things.
- Social Learning: as their type's name indicates, social learners are socialist and prefer to learn in groups rather than on their own. Their learning technique involves generating new ideas by brainstorming with others.
- Solitary Learning: Solitary learners are opposite in nature to social learners. They prefer to be more independent and spending much time on their own.

*3.2. Learning Style Inventories (LSIs) -*

As mentioned before, to identify the students learning types to maximize their academic confidence and provide information in terms of what they are comfortable with, teachers are conducting learning style inventories and then examining the inventories results of the students' learning styles to choose the appropriate teaching methodologies and styles [16]. There are many learning styles' inventories. Below is a list of the popular ones:

- *Fleming's VARK Questionnaire:*
  It is one of the common and well-known used questionnaires. VARK denotes Visual, Auditory, Reading/writing, and Kinesthetic, which reflect the four different learning styles that Fleming identified. To help people know their individual styles, a questionnaire was designed in 1987 [12]. Since then, the questionnaire has become very popular and widely used today. The following subsection (3.3) explains this inventory in detail.

- *Kolb's Learning Style Inventory:*
  Kolb's inventory is designed based on experiential learning theory to help individuals identify how they learn from experience. Kolb's learning styles break up into four levels: Diverge, Assimilator, Accommodator, and Converge [13].

- *Jackson's Learning Styles Profiler:*
  This Profiler (LSP) is designed based on Chris J. Jackson's hybrid learning model in personality. Jackson argues that there is a common biological basis for positive and negative outcomes within the workplace, education, and the general community. Jackson's profiler is commonly used in business settings as it is designed to assess how people learn at work [14].

- *Online Learning Style Quizzes:*
  Many free informal online questionnaires are also available online. However, these questionnaires have never been validated and are just to gain some knowledge about the users learning preferences. So, it is not recommended to use such questionnaires.

3.3 *VARK* Model -

In Fleming's or VARK model, four key types of learners are identified, which are Visual (preferred Graphic displays such as chart and diagrams), Auditory (learn best by hearing information and conducting discussions), Reading/writing (preferred reading textbooks), and Kinesthetic learners (preferred movement, experiments, and Hands-on experience are important). Fleming also developed a self-report inventory with many questions about choosing the best match to their preferred approach or learning style [12]. For example, suppose the following question:

To learn something new on a computer, I would:

   a) Follow the diagrams in the user manual.
   b) Read the written instructions in the user manual.
   c) Listen to people who already do that task.
   d) Start using it and then learn by trial and error.

If the respondent chooses the first answer, then he/she might be a visual learner. However, if the respondent chooses the second answer (would prefer to read written instructions), then he/she is likely a reading/writing learner. On the other hand, if somebody prefers to listen to someone explaining doing a specific task, then he is an auditory learner. Finally, those who prefer to carry out physical activities are kinesthetic learners. This questionnaire will be utilized in our research as it is very suitable for students and educators.

4. Methodology

The methodology employed in this study requires a clear understanding of the tradeoffs inherent in this domain. In fact, the key challenge is to classify learners according to their distinguishing features (learning styles), taking into account that learners may have a mix of learning styles with the probability of having no dominant style of learning. As such, our general approach builds upon the regression analysis to provide a probabilistic approach for inferring the preferred learning styles. Because of this domain's maturity in general, it is important to ground our approach with a robust experimental evaluation. To this end, we constructed multiple prediction models using five machine learning algorithms. The dataset used in these experimental tests was collected using the VARK's inventory questionnaire from a sample of 72 students.

To demonstrate that our approach is more accurate than classification, we develop classification models for inferring the learning style label using the same set of machine learning algorithms. The models are evaluated using recall, precision, accuracy, F1-score, and Area Under Curve (AUC). Based on the obtained results, one can conclude that regression algorithms are more accurate and representative for predicting learning styles' probabilities.

## 5. Materials and Methods

The Materials and Methods include a short description of the data collection method, details of the data preprocessing process, and the machine learning techniques and methodology used to predict learning styles.

*5.1 Data collection*

To conduct our study, a sample of 72 students was randomly selected from the higher education institutes. The sample data was collected using VARK's inventory questionnaire, where four different learning styles are identified: Visual (V), Auditory (A), Reading/Writing (R), and Kinesthetic (K). The questionnaire consists of 16 different questions that deal with the way(s) in which students like to learn or prefer to deliver. The questions are based on situations where there are choices and decisions about how those might happen. To easily collect the students' responses, we developed an online version using Microsoft Forms. The responses are then imported as an Excel file where each answer is represented as a vector of binary values denoted as <A, V, K, R>. The data is then preprocessed (as described in Section 5.2) to be eligible for the employed machine learning algorithms. Here, we divided the whole dataset into an array of four matrices, a matrix for each learning style. Each matrix consists of 16 columns demonstrating the presence or absence of a learning style and 5 output columns (4 columns represent the learning styles' probabilities, and the last column represents the selected learning style label).

Figure 1 shows the probability distribution for each learning style using boxplots. The horizontal line inside the boxplot represents the median of probabilities, while the x symbol inside each box represents the mean of probabilities. We can observe that all learning styles have a relatively similar distribution. For instance, learning style (A) has the largest mean and median values with narrower distribution than other learning styles, suggesting that all students have the same learning style probabilities (A). Also, we can notice that most students favored learning style (A). Based on visual observation, we can see that the mean of probabilities for learning style (A) is significantly different from that of (V) and (R) learning styles.

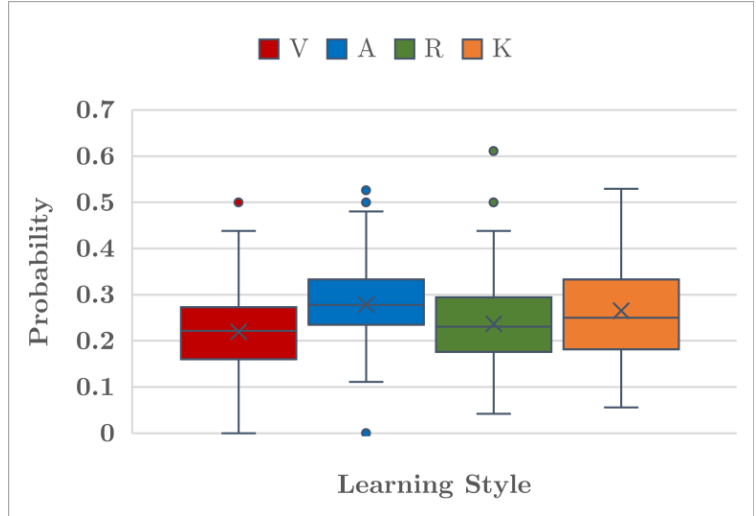
Figure 1. Boxplots of Probabilities for all learning styles

Figure 2 shows a bar plot of all learning styles based on the actual responses provided by students. We can observe that learning style (A) is the dominant one of 22 selection followed by learning style (K) with 19 selection, then learning style (V) with 17 selection and finally learning style (R) with 13 selection.

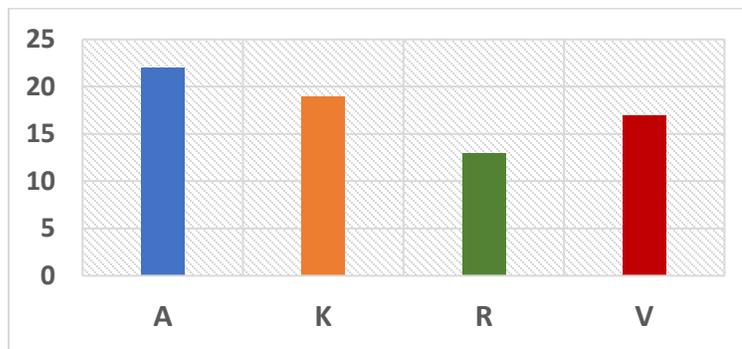
Figure 2. Bar plot of class labels

Figure 3 shows the bar plot of each learning style across all questions. In other words, for each question, we record the frequency of each learning style. From the Figure, many points can be noticed: 1) Although learning style (A) was the dominant label for students' styles, surprisingly, we can notice that learning style (A) is the most frequent style for two questions only (Q1, Q3). This happens because we take individual count for each question, not probabilities across student responses. 2) The learning style (V) was the top selected response for also two questions (Q10 and Q14), whereas learning style (K) was the top selected response in seven questions (Q4, Q5, Q7, Q8, Q11, Q13, and Q16). 3) The learning style (R) was the top selected response on five questions (Q2, Q6, Q9, Q12, and Q15). 4) For some questions, there is a convergence between top frequencies, such as on Q8 and Q13, where the difference between the frequency of Learning style (A) and learning style (K) is neglectable.

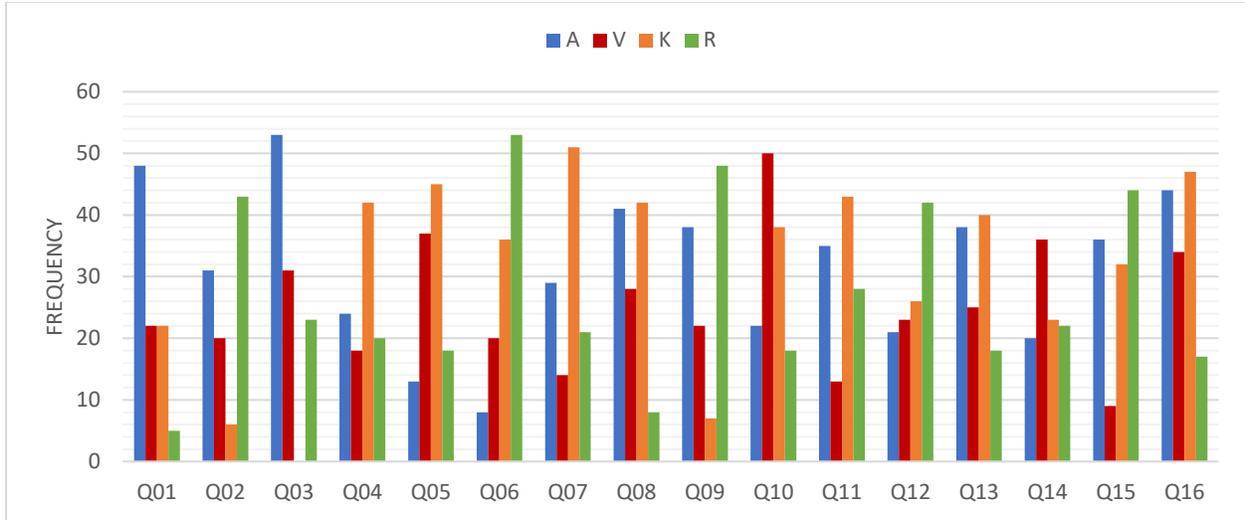
Figure 3. Bar plot of learning styles distribution across all questions

*5.2 Data preprocessing*

The collected dataset has been first preprocessed to be eligible for the used machine learning algorithms. To this end, the original dataset is described by multiple rows and columns, where each row represents student responses, and the columns represent questions. Each response consists of a list of one or more styles selected from the complete response labels {A, V, K, R}. In other words, each student might provide multiple styles for the response to each question. To facilitate processing the data, we represent each answer as a vector of binary values denoted as <A, V, K, R>. For example, the vector <1, 0, 1, 0> means that the student responded with A and K learning styles. The processed data has been manipulated to look like the dataset shown in Table 1, where the last five columns are considered output. Four of them are considered numeric, which are denoted by 'Prob of', which is the probability of each learning style that will be served as output for regression models. Whereas the last column is the actually selected learning style label, given based on the maximal probability, which served as output for classification models.

Table 1. Processed Dataset

| ID | Q1 | Q2 | Q3 | ... | Q16 | Prob of A | Prob of V | Prob of K | Prob of R | Learning Style |
|---|---|---|---|---|---|---|---|---|---|---|
| S1 | <1,1,0,1> | <1,0,1,1> | <1,0,0,0> | ... | <0,0,1,1> | 0.36 | 0.24 | 0.25 | 0.15 | A |
| S2 | <0,0,0,1> | <0,1,0,1> | <1,0,0,1> | ... | <0,1,0,0> | 0.18 | 0.44 | 0.21 | 0.17 | V |
| S3 | <0,0,0,1> | <0,1,0,0> | <0,1,1,1> | ... | <1,1,0,0> | 0.14 | 0.25 | 0.19 | 0.42 | R |
|  |  |  |  | ... |  |  |  |  |  |  |
| N | <0,1,0,0> | <0,0,0,1> | <0,0,0,1> | ... | <0,1,0,1> | 0.42 | 0.49 | 0.07 | 0.02 | V |

Since each cell contains a binary vector representing student response, we divided the whole dataset into an array of four matrices, a matrix for each learning style. Specifically, each matrix has the same number of rows and columns, but each question cell represents the corresponding value for that learning style. For example, the matrix for learning style R will look as shown in Table 2. All matrices share the same set of numeric and label outputs.

Table 2. Learning Style (R) Matrix

| ID | Q1 | Q2 | Q3 | ...... | Q16 | Prob of A | Prob of V | Prob of K | Prob of R | Learning Style |
|----|----|----|----|--------|-----|-----------|-----------|-----------|-----------|----------------|
| S1 | 1  | 1  | 0  | ...... | 1   | 0.36      | 0.24      | 0.25      | 0.15      | A              |
| S2 | 1  | 1  | 1  | ...... | 0   | 0.18      | 0.44      | 0.21      | 0.17      | V              |
| S3 | 1  | 0  | 1  | ...... | 0   | 0.14      | 0.25      | 0.19      | 0.42      | R              |
|    |    |    |    | ...... |     |           |           |           |           |                |
| N  | 0  | 1  | 1  | ...... | 1   | 0.42      | 0.49      | 0.07      | 0.02      | V              |

## 5.3 Constructed Models

Two kinds of prediction models have been constructed in this paper, one for regression and one for classification. We used probabilities as output for regression models and learning style labels as output for classification. Then we aggregate results for each learning style as prediction and classification. The same set of machine learning algorithms have been used for both regression and classification. These algorithms are Multi-Layers Perceptron Neural Network (NN), Support Vector Machine (SVM), Decision Tree (DT), Random Forest (RF), and K-Nearest Neighbors (kNN).

k-Nearest Neighbor (kNN) is a supervised machine learning algorithm based on the idea of predicting by similarity. Specifically, the kNN uses a distance measure such as Euclidean distance to retrieve, for each new observation, the nearest observations from the training dataset. The final output is then predicted from the outputs of the set of selected observations. Many control parameters impact the performance of kNN, such as feature selection, voting mechanism, feature weighting, number of selected observations ($k$), and type of distance measure. It is important to note that a smaller value of $k$ can be noisy and subject to the effect of an outlier, whereas the large value of $k$ can smooth the final decision [28]. In this paper, we used the common rule of thumb that states $k = \sqrt{n}$, where $n$ is the number of training observations.

SVM is basically build an optimal hyperplane that can separate data with maximum margin. The margin is defined as the maximal width of the slab parallel to the hyperplane with no interior data points. The optimal hyperplane generation depends on the choice of kernel functions such as Gaussian, Polynomial, and Radial Basis Function (RBF). Both Gaussian and Radial Basis function kernels can benefit hyperplane generation because they support the locality of training data, which means that the data can be efficiently separated [29]. In this paper, we only used the RBF kernel.

DT is a tree-like method that uses iterative partitioning to construct the tree. The algorithm uses probabilistic measures such as entropy, information gain, and Gini metrics to decide which optimal features should be used to separate data at each decision node. In each step, more coherent data are grouped based on the decision node. The algorithm also uses a pruning algorithm to remove unwanted sub-branches that do not contribute significantly to the decision process. In this paper, the C4.5 algorithm has been used because it can create more generalized trees and not fall into overfitting, and it can also handle incomplete data very well [30].

Neural Network used in this paper is a kind of Multi-Layers Perceptron with one input layer, at least one hidden layer, and one output layer. Each neuron of the input layer represents an input vector. A nonlinear activation function is usually used in the hidden layer neurons, whereas a linear activation function is usually used in the output layer. The number of neurons in the hidden layer varies based on the number of input neurons and the type of training algorithm used. One popular training algorithm is the backpropagation algorithm, a type of gradient descent algorithm [31].

Random Forest is an ensemble learning algorithm constructed from a set of decision trees using the Bagging algorithm. RF adds additional randomness to the model while growing the trees. Instead of searching for the most important feature while splitting a node, it searches for the best feature among a random subset of features. This results in a wide diversity that generally results in a better model. Therefore, in a random forest, only a random subset of the features is considered by the algorithm for splitting a node. You can even make trees more random by additionally using random thresholds for each feature rather than searching for the best possible thresholds (like a normal decision tree does) [32].

$$MAE = \frac{\sum_{i=1}^{n}|x_i - \hat{x}_i|}{n} \tag{1}$$

$$MdAE_i = median_{\forall i}(|x_i - \hat{x}_i|) \tag{2}$$

$$RMSE = \sqrt{\frac{1}{n}\sum_{i=1}^{n}|x_i - \hat{x}_i|^2} \tag{3}$$

Where $x_i$ and $\hat{x}_i$ are the actual and estimated probability of learning style.

The constructed model attempts to predict the probability of each learning style to identify the most favoured styles. In this case the output of prediction would be in this format: <A=0.3, V=0.22, K=0.08, R=0.4>. Then a threshold can be specified to select the most favoured learning styles. To accomplish that, we compute the distance between the top learning style and remaining learning styles, then any learning style that falls within the distance given by the threshold is selected as nominated learning style. In the example above, if the threshold equals 0.2, then the selected learning style is {R, A, V}. If the threshold is 0.1, then the selected learning style set is {R, A}. We recommend the threshold value to be not very small that would ignore some important learning style or too large involving all learning styles.

In the first experiment, these algorithms have been applied for each matrix to predict each learning style's probability as a regression problem. We record the Mean of Absolute Errors (MAE) as shown in equation 1, Median of absolute Errors (MdAE) as shown in equation 2, Root Mean of Squared Errors (RMSE) as shown in equation 3. Then, we aggregate MAE, MdAE, and RMSE using the average aggregation method for each learning style probability. In the second experiment, we used the same set of machine learning algorithms on each matrix as a classification to predict only the Learning style label. The models have been evaluated using recall as shown in equation 4, precision as shown in equation 5, F1-score as shown in equation 6, accuracy as shown in equation 7, and Area Under Curve (AUC). The aggregate evaluation values for each learning style have been recorded.

Recall is the proportion of the observations that are correctly predicted for each class label.

$$Recall = \frac{TP}{TP + FN} \tag{4}$$

Precision is the proportion of the observations that are tested correctly for each class label.

$$Precision = \frac{TP}{TP + FP} \tag{5}$$

F1 is the balance measure between precision and recall

$$F1 = 2 \times \frac{(Recall \times Precision)}{Recall + Precision} \tag{6}$$

Accuracy is the overall indicator of the classification performance.

$$Accuracy = \frac{TP + TN}{TP + TN + FP + FN} \tag{7}$$

Where TP (True Positive) is the number of positive observations that are predicted as such. TN (True Negative) is the number of negative observations that are predicted as such. FP (False Positive) is the number of negative observations that are predicted as positive. FN (False Negative) is the number of positive observations that are predicted as negative.

## 6. Results and Discussion

This section presents the results of applying both regression and classification models to the collected dataset. We first present the results obtained from the regression models, and then we will focus on the classification models.

*6.1 Regression models results*

For the regression task, we used the probabilities of each learning style as output. As mentioned in the methodology section, we have constructed an array of four binary matrices, where each matrix represents the presence or absence of a learning style, and the output is the probabilities of all learning styles. For each matrix, we applied five popular machine learning algorithms as regression to predict the probability of each learning style, which are NN, SVM, DT, RF, and *k*NN. Then we aggregate results for each learning style from the four matrices. The predicted probabilities are compared to the actual probabilities using four performance metrics, as shown in Table 3. The values with boldface and underline represent the most accurate results. This Table shows that RF is almost the superior one for predicting all learning styles across all performance metrics. For learning style (A), we found that RF and SVM work well for predicting the probabilities of (A) for all students. For Learning (V) and (R), we found that RF is the dominant model across all performance metrics.

Table 3. Accuracy Results of all Regression models

| Metric | A | | | | | V | | | | | K | | | | | R | | | | | All | | | | |
| --- | --- | --- | --- | --- | --- | --- | --- | --- | --- | --- | --- | --- | --- | --- | --- | --- | --- | --- | --- | --- | --- | --- | --- | --- | --- |
| | NN | SVM | kNN | DT | RF | NN | SVM | kNN | DT | RF | NN | SVM | kNN | DT | RF | NN | SVM | kNN | DT | RF | NN | SVM | kNN | DT | RF |
| MAE | 0.0849 | 0.0614 | 0.0640 | 0.0653 | 0.0612 | 0.0962 | 0.0622 | 0.0713 | 0.0653 | 0.0614 | 0.0923 | 0.0737 | 0.0793 | 0.0731 | 0.0730 | 0.0831 | 0.0688 | 0.0701 | 0.0728 | 0.0685 | 0.0713 | 0.0532 | 0.0569 | 0.0553 | 0.0528 |
| MdAE | 0.0671 | 0.0405 | 0.0405 | 0.0476 | 0.0451 | 0.0882 | 0.0523 | 0.0573 | 0.0580 | 0.0520 | 0.0812 | 0.0615 | 0.0608 | 0.0564 | 0.0594 | 0.0732 | 0.0561 | 0.0560 | 0.0595 | 0.0541 | 0.0686 | 0.0519 | 0.0497 | 0.0517 | 0.0507 |
| RMSE | 0.1102 | 0.0835 | 0.0864 | 0.0853 | 0.0843 | 0.1184 | 0.0786 | 0.0897 | 0.0816 | 0.0773 | 0.1115 | 0.0930 | 0.1043 | 0.0941 | 0.0912 | 0.1038 | 0.0882 | 0.0909 | 0.0936 | 0.0867 | 0.0762 | 0.0594 | 0.0631 | 0.0602 | 0.0580 |

The Wilcoxon significance tests between each pair of models based on absolute errors over each learning style are presented in Table 4. The results show that predictions produced by NN are almost different from those generated by other prediction models. However, the accuracy of NN in table 4 was significantly different from other models but not necessarily superior. On the other hand, we did not have any significant difference between each pair of models, which means that all prediction models, except NN, produce relatively similar predictions. From these results, we can conclude that NN is the only model that can generate different predictions than others, while all remaining models behave similarly overall learning styles. Interval plots confirm these results in Figure 4.

Table 4. Wilcoxon statistical significant test of absolute residuals between each pair of models

| Model 1 | Model 2 | A | V | K | R | All |
|---|---|---|---|---|---|---|
| NN | SVM | 0.036 | 0.002 | 0.06 | 0.02 | 0.003 |
| NN | kNN | 0.061 | 0.036 | 0.1 | 0.032 | 0.006 |
| NN | DT | 0.13 | 0.01 | 0.04 | 0.02 | 0.006 |
| NN | RF | 0.036 | 0.0034 | 0.05 | 0.02 | 0.0001 |
| SVM | kNN | 0.82 | 0.41 | 0.93 | 0.9 | 0.76 |
| SVM | DT | 0.54 | 0.71 | 0.84 | 0.96 | 0.80 |
| SVM | RF | 0.84 | 0.97 | 1.00 | 0.99 | 0.99 |
| kNN | DT | 0.66 | 0.6 | 0.74 | 0.85 | 0.89 |
| kNN | RF | 0.71 | 0.41 | 0.94 | 0.89 | 0.7 |
| DT | RF | 0.45 | 0.66 | 0.83 | 0.96 | 0.76 |

Figure 4 shows the interval plots of absolute errors for all prediction models over each learning style. It is clear that NN produces significantly different predictions than other models of overall learning styles, as shown in Table 4. Despite that, the NN produces bad results than other models. The remaining models behave similarly with no significant differences between their predictions, which means that any one of them can perform the job. But, we recommend using the most accurate one which is RF.

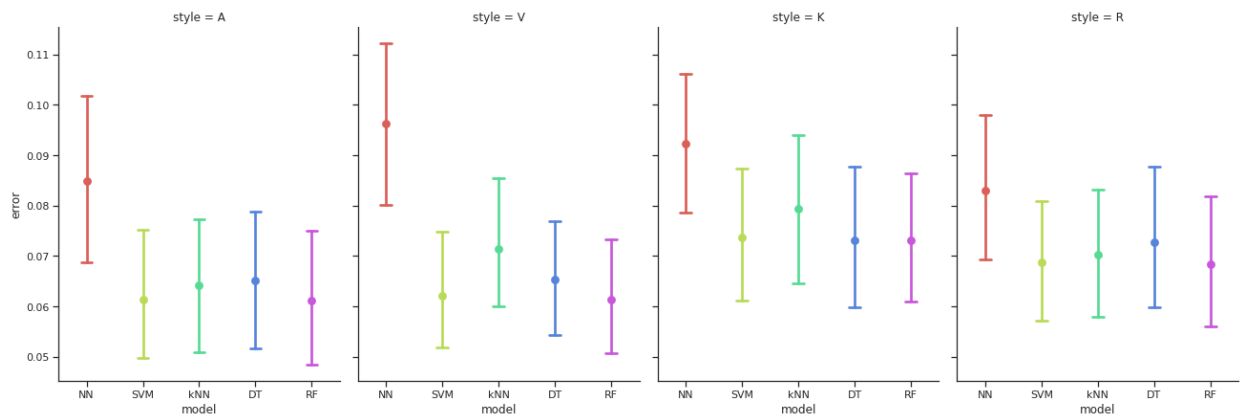

Figure 4. Comparison between prediction models for each learning style, using interval plots

### 6.2 Classification models results

The second experiment is designed to predict the correct learning style for each student using classification algorithms. We used the same machine learning methods mentioned above, but this time as a classification approach. For each matrix, we neglected numeric probabilities and used the dominant label, and then we apply classification algorithms over each matrix. The classification accuracy is presented over each learning style matrix using the popular evaluation measures such as precision, recall, F1-score, accuracy, and Area Under Curve (AUC). Tables 5 to 8 show the accuracy results for all classification models over each learning style matrix. The first imprecision from these tables is that none of the classification models is predictive with high accuracy.

Table 5 shows the evaluation results over the learning style (A) matrix. We have found that RF is the most superior one across all evaluation metrics. The overall results are not encouraging, suggesting that none of the models can produce highly accurate predictions. The ROC plots of all models are not encouraging, as shown in Figure 5. However, amongst them, we can notice that RF is the most predictive one.

Table 5. Evaluation results over the matrix of learning Style A

| Model | Precision | Recall | F1-Score | Accuracy | AUC |
|---|---|---|---|---|---|
| SVM | 0.368 | 0.421 | 0.377 | 0.479 | 0.654 |
| NN | 0.405 | 0.403 | 0.402 | 0.423 | 0.695 |
| RF | 0.505 | 0.506 | 0.503 | 0.535 | 0.713 |
| DT | 0.413 | 0.417 | 0.408 | 0.451 | 0.632 |
| KNN | 0.423 | 0.422 | 0.398 | 0.451 | 0.682 |

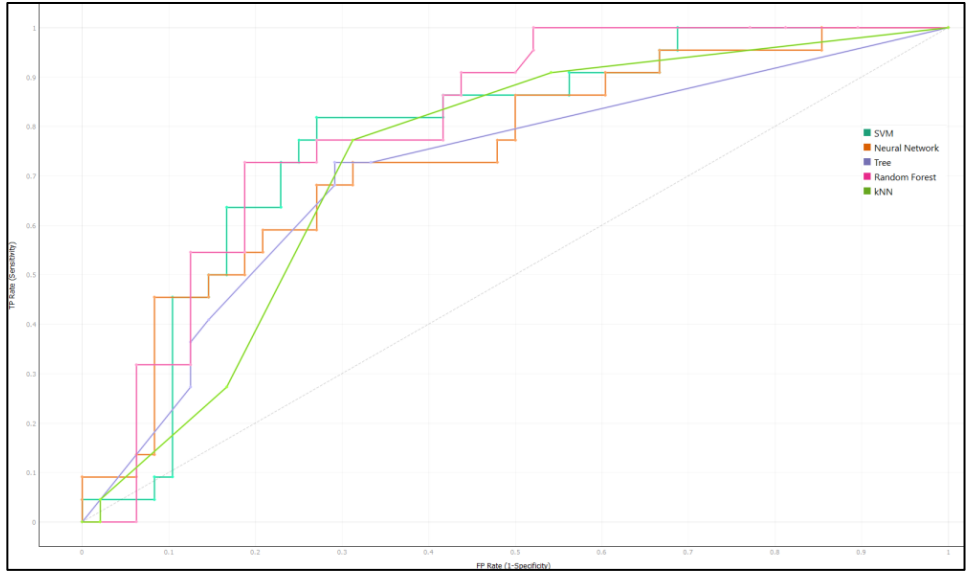

Figure 5. ROC plot of all models over the matrix of learning style A.

The results over the learning style matrix (K) are very bad in general. We could not find a stable conclusion for which model is more predictive and superior. Both DT and SVM are considered the most predictive across all evaluation measures. The ROC plots in Figure 6 are worse than that of Figure 5. From the plots, we can observe that SVM is the superior one, and RF is the worst one.

Table 6. Evaluation results over the matrix of learning Style K

|  | Precision | Recall | F1-Score | Accuracy | AUC |
|---|---|---|---|---|---|
| SVM | 0.371 | 0.444 | 0.401 | 0.491 | 0.643 |
| NN | 0.295 | 0.299 | 0.321 | 0.324 | 0.588 |
| RF | 0.340 | 0.356 | 0.338 | 0.380 | 0.545 |
| DT | 0.430 | 0.409 | 0.404 | 0.437 | 0.610 |
| KNN | 0.279 | 0.276 | 0.252 | 0.310 | 0.576 |

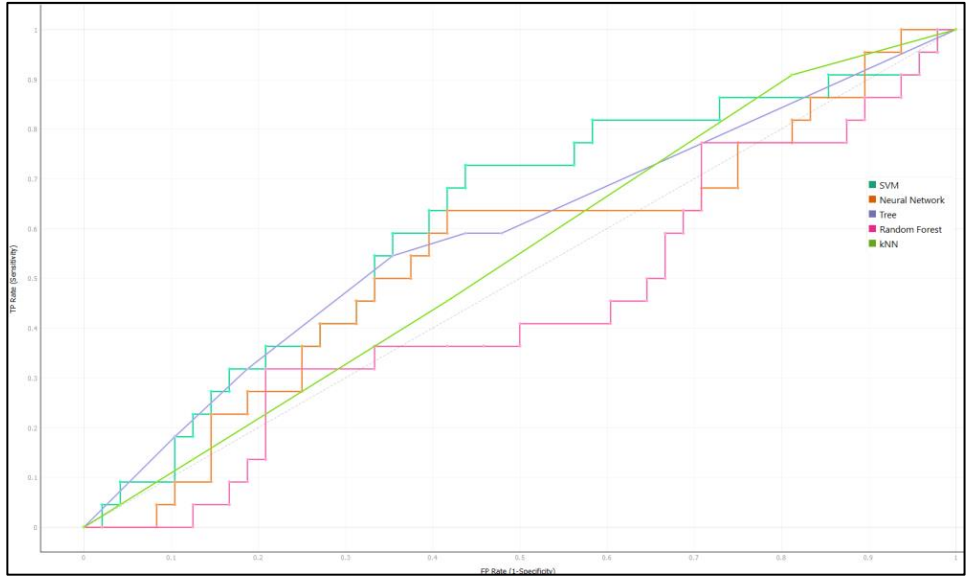
*Figure 6. ROC plot of all models over the matrix of learning style K.*

The results obtained over the learning style (R) matrix are more stable than the learning style matrix (K). Here, the SVM is the most stable, with high precision, recall, and accuracy but with less AUC than KNN. Interestingly, all models' AUC values are at the average level, confirmed by ROC plots in Figure 7. It is clear from ROC plots that all models are under the threshold line, which means that all models are not predictive over the (R) matrix.

Table 7. Evaluation results over the matrix of learning Style R

|     | Precision | Recall | F1-Score | Accuracy | AUC   |
| --- | --------- | ------ | -------- | -------- | ----- |
| SVM | 0.331     | 0.303  | 0.309    | 0.310    | 0.511 |
| NN  | 0.298     | 0.286  | 0.290    | 0.282    | 0.530 |
| RF  | 0.294     | 0.274  | 0.280    | 0.268    | 0.500 |
| DT  | 0.288     | 0.280  | 0.281    | 0.282    | 0.512 |
| KNN | 0.281     | 0.295  | 0.279    | 0.282    | 0.553 |

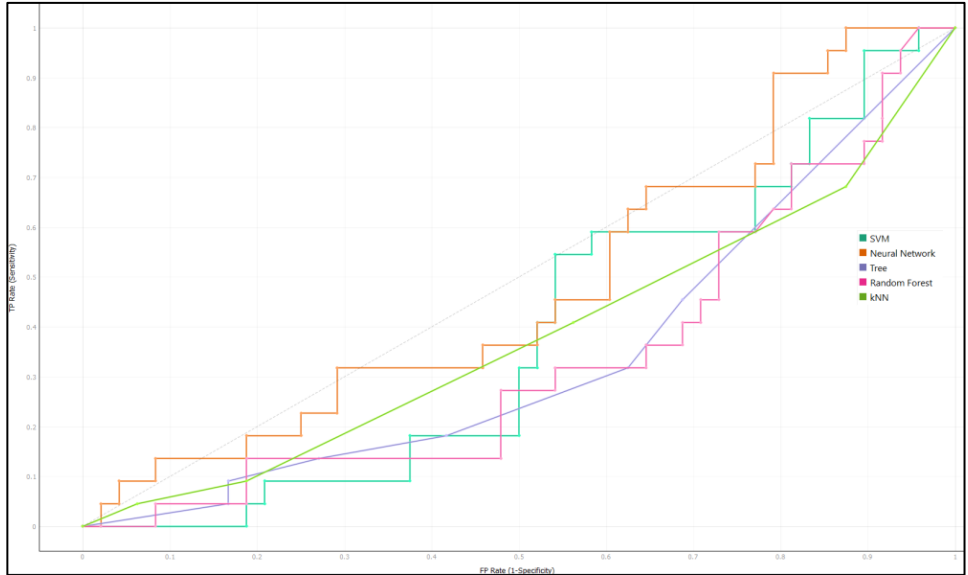
*Figure 7. ROC plot of all models over the matrix of learning style R.*

Finally, the same trend can be seen over the learning style matrix (V), where the NN and SVM are the most predictive ones. The ROC plots in Figure 8 are also not encouraging, even though the AUC values are generally reasonable for all models.

Table 8. Evaluation results over the matrix of learning style V

|     | Precision | Recall | F1-Score | Accuracy | AUC   |
|-----|-----------|--------|----------|----------|-------|
| SVM | 0.384     | 0.339  | 0.328    | 0.352    | 0.629 |
| NN  | 0.368     | 0.376  | 0.371    | 0.366    | 0.586 |
| RF  | 0.301     | 0.299  | 0.300    | 0.296    | 0.565 |
| DT  | 0.241     | 0.241  | 0.237    | 0.254    | 0.535 |
| KNN | 0.308     | 0.288  | 0.292    | 0.296    | 0.610 |

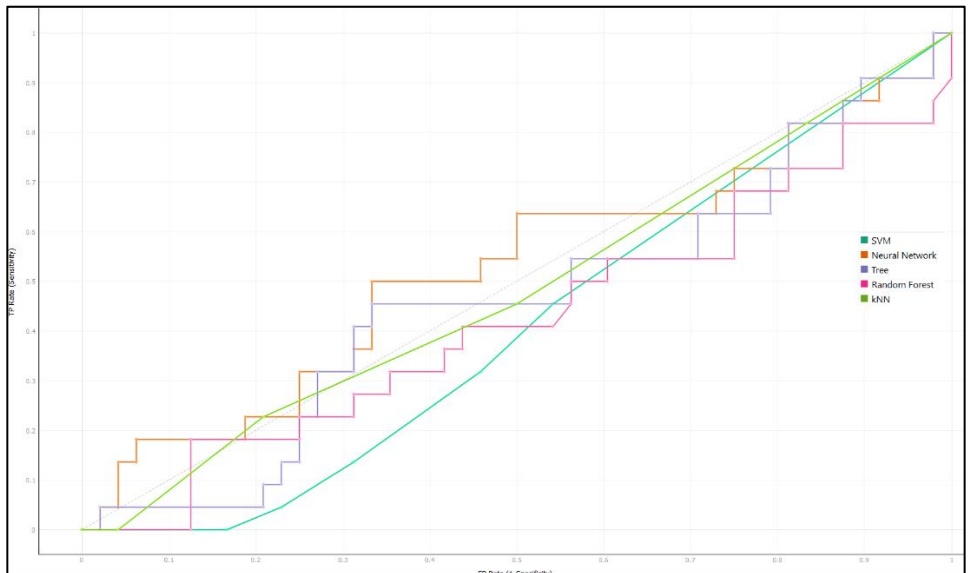
*Figure 8. ROC plot of all models over the matrix of learning style V.*

From the above results (both regression and classification), we can conclude the predicting optimal learning style for students as classification is not accurate as predicting the learning style as probabilities. Therefore, we found that using regression algorithms to predict the learning styles probabilities are more accurate and representative for real-world scenarios where students might choose multiple learning styles but with different probabilities.

Finally, it is important to mention here that several limitations are apparent in our study's last part. The size sample was relatively small, and they study at the same university what might have influenced the study results. The results would be more precise if the sample size is larger, and it took from different universities.

7. Conclusion

Students might find that understanding their own learning preferences are helpful. This is supported by recognizing the students' learning styles and approved by many studies that found the use of learning styles in conjunction with other learning methods enhances academic achievements or, at the very least, makes studying more enjoyable. This study is a mixed-method approach that aims to predict the learning styles for learners with mixed styles (with probability). To this end, theories and strategies have been investigated that identify the students' features according to their learning styles. Then the regression analysis was utilized to provide a probabilistic approach for predicting the preferred learning styles. Here, five machine learning algorithms were applied as regression to predict the probability of learning styles, which are Multi-Layers Perceptron Neural Network (NN), Support Vector Machine (SVM), Decision Tree (DT), Random Forest (RF), and K-Nearest Neighbors. A sample of 72 students was randomly selected to conduct our study. The sample data was collected using VARK's inventory questionnaire with 16 different questions to identify four different learning styles: Visual (V), Auditory (A), Reading/Writing (R), and Kinesthetic (K). Results showed that the RF algorithm was the superior one for predicting the probabilities of all learning styles.

To examine the effectiveness of our approach, the same set of machine learning algorithms were used to develop classification models for predicting the learning style label. We aimed to compare our finding with the classification-based approach. We observed from the results that accuracies of all classification models are relatively low. The RF showed the best accuracy with 0.53 to predict learning style (A). Moreover, the overall results are not encouraging, suggesting that none of the models can produce highly accurate predictions. So, we conclude that regression algorithms are more accurate and representative for predicting learning styles' probabilities.

Acknowledgement

The authors are grateful to the Applied Science Private University, Amman-Jordan, for the full financial support granted to cover the publication fee of this research article.